\documentclass[letterpaper, 10 pt, conference]{ieeeconf}  

\IEEEoverridecommandlockouts                              

\usepackage{graphicx} 
\usepackage{mathptmx} 
\usepackage{times} 
\usepackage{amsmath} 
\usepackage{amssymb}  
\usepackage{textcomp}
\usepackage{booktabs}
\usepackage{multirow}
\usepackage{balance}
\usepackage{soul,color}
\usepackage{comment}
\usepackage{array}

\newcommand {\etal}{\emph{et al.}}
\newcommand {\eg}{\emph{e.g.}}
\newcommand {\ie}{\emph{i.e.}}

\title{\LARGE \bf
Neural Memory Networks for Seizure Type Classification 
}

\author{David Ahmedt-Aristizabal$^{1,2}$, 
Tharindu Fernando$^{2}$, 
Simon Denman$^{2}$,\\
Lars Petersson$^{1}$,
Matthew J. Aburn$^{3}$,
Clinton Fookes $^{2}$ 
\thanks{$^{1}$ CSIRO, DATA61, Canberra, Australia.%
}%
\thanks{$^{2}$ Image and Video Research Laboratory, SAIVT, Queensland University of Technology, Brisbane, Australia.%
}%
\thanks{$^{3}$ QIMR Berghofer Medical Research Institute, Brisbane, Australia..%
}%
}

\begin{document}

\maketitle
\thispagestyle{empty}
\pagestyle{empty}

\begin{abstract}
Classification of seizure type is a key step in the clinical process for evaluating an individual who presents with seizures. It determines the course of clinical diagnosis and treatment, and its impact stretches beyond the clinical domain to epilepsy research and the development of novel therapies.
Automated identification of seizure type may facilitate understanding of the disease, and seizure detection and prediction have been the focus of recent research that has sought to exploit the benefits of machine learning and deep learning architectures. Nevertheless, there is not yet a definitive solution for automating the classification of seizure type, a task that must currently be performed by an expert epileptologist.
Inspired by recent advances in neural memory networks (NMNs), we introduce a novel approach for the classification of seizure type using electrophysiological data. We first explore the performance of traditional deep learning techniques which use convolutional and recurrent neural networks, and enhance these architectures by using external memory modules with trainable neural plasticity. 
We show that our model achieves a state-of-the-art weighted F1 score of 0.945 for seizure type classification on the TUH EEG Seizure Corpus with the IBM TUSZ preprocessed data. 
This work highlights the potential of neural memory networks to support the field of epilepsy research, along with biomedical research and signal analysis more broadly.
\end{abstract}

\section{INTRODUCTION}

Epilepsy is one of the most prevalent neurological conditions and people with epilepsy have recurrent seizures. Separating individual seizures into different types helps guide antiepileptic therapies~\cite{fisher2017operational}.
Classification of seizures serves many purposes. It is informative of 
the potential triggers for a patient's seizures, the risks of comorbidities including intellectual disability,
learning difficulties, mortality risk such as sudden unexpected death from epilepsy, and psychiatric features such as autism spectrum disorder\cite{fisher2017operational}. 

Together with observation of clinical signs, electroencephalography (EEG) plays a major role in seizure type evaluation and automating this process can support clinical evaluation.
Recent advances in artificial intelligence and deep learning have demonstrated high success in other healthcare applications using brain signals~\cite{faust2018deep}.
However, application of these architectures within neuroscience and specifically to the processing of EEG recordings for epilepsy research, have been limited to date~\cite{ahmedt2017automated,craik2019deep}. 
Current deep learning approaches have mostly focused on the goals of seizure detection~\cite{zabihi2019patient,ahmedt2018deep} and seizure onset prediction~\cite{kuhlmann2018seizure}; and deep convolutional neural networks (CNN) and recurrent neural networks (RNN) have been the most common architectures proposed to capture patterns during seizures~\cite{zhang2019epilepsy,thodoroff2016learning,golmohammadi2017gated}.
Nevertheless, the automated capability to discriminate among seizure types (\eg~focal or generalized seizures) is a challenging and largely underdeveloped field due to both a lack of datasets and the highly complex nature of the task.
%
A significant public data resource, the TUH EEG Corpus~\cite{shah2018temple}, has recently become available for epilepsy research, creating a unique opportunity to evaluate deep learning techniques.
To date, a limited set of methods have been applied to this challenging corpus in full for the task of seizure classification~\cite{roy2019machine,asif2019seizurenet,sriraam2019convolutional}, although some researchers have used a small number of data sample from selected seizure types as input for their models~\cite{saputro2019seizure,saputro2018tonic,song2019dynamic}.

Motivated by the tremendous success of neural memory networks to precisely store and retrieve relevant information~\cite{munkhdalai2017neural,fernando2018task,fernando2018tree},
we propose a novel approach based on long-term memory modules to identify and exploit relationships across the entire EEG data set for seizure events.
We capture the variability, both intrasubject (seizures of the same patient) and intersubject (seizures across patients), for each epilepsy type in this long-term relationship.
One of the main limitations of using traditional recurrent neural networks such as Long Short Term Memory (LSTM)~\cite{greff2016lstm} or Gated Recurrent Unit (GRU)~\cite{cho2014learning} layers with seizure recordings is that they focus more on the recent history and previous memories are lost after updates~\cite{chen2016enhancing}, \ie~they consider dependencies only within a given input sequence.
To address this limitation, we need to extract and store events over time, and this is possible with an external memory bank. In this scenario, a framework with an external memory should also learn when to store an event, as well as when to recall it for use in the future~\cite{fernando2018tree}. With the help of the external memory, the network no longer needs to squeeze all useful past information into the state variable (the cell state that saves information from the past) of the LSTM or GRU.
We also adopt the concept of synaptic plasticity, which emulates the biological process of the same name to enable efficient lifelong learning, and to enhance the attention based knowledge retrieval mechanisms used in memory networks~\cite{miconi2018differentiable}. 
The plastic neural memory exploits both static and dynamic connection weights in the memory read, write and output generation procedures (\ie~a connection between any two neurons has both a fixed component and a plastic component)~\cite{tharindu2019TNNLS}. 

In this research, we perform cross-patient seizure type classification, with an application of supporting the analysis of scalp EEG seizure recordings where epileptologists are not available.
We first explore the feasibility of adapting deep learning algorithms that have shown promising results for seizure detection for the specific evaluation of seizure type classification.
Then, we introduce our framework based on memory networks and trainable neural plasticity~\cite{miconi2018differentiable}, which is a mechanism for knowledge discovery, \ie~a dynamic strategy to read and write relevant information to capture temporal relationships. We expand on the work introduced for anomaly detection~\cite{tharindu2019TNNLS} to demonstrate the potential of our architecture for the complex task of multi-class classification of seizures, and compare the results with previously published baseline methods.

Our technical contributions are summarized as follows:
\vspace{-2pt}
\begin{enumerate}
\item This study presents baseline results that compare several traditional deep learning algorithms proposed for seizure detection optimised and evaluated for the task of classifying seizure types.
\item We propose a robust approach based on neural memory networks which outperforms state-of-the-art methods for seizure type classification on the TUH EEG Seizure corpus~\cite{shah2018temple}. 
\item We introduce the first application of memory modules, which are capable of mapping long-term relationships, to the field of epilepsy research and demonstrate how they can provide a clear separation between classes using the extracted memory embeddings.
\end{enumerate}
\vspace{-2pt}

\section{MATERIALS AND METHODS}
\begin{figure*}[!h]
\begin{center}
\includegraphics[width=0.77\linewidth]{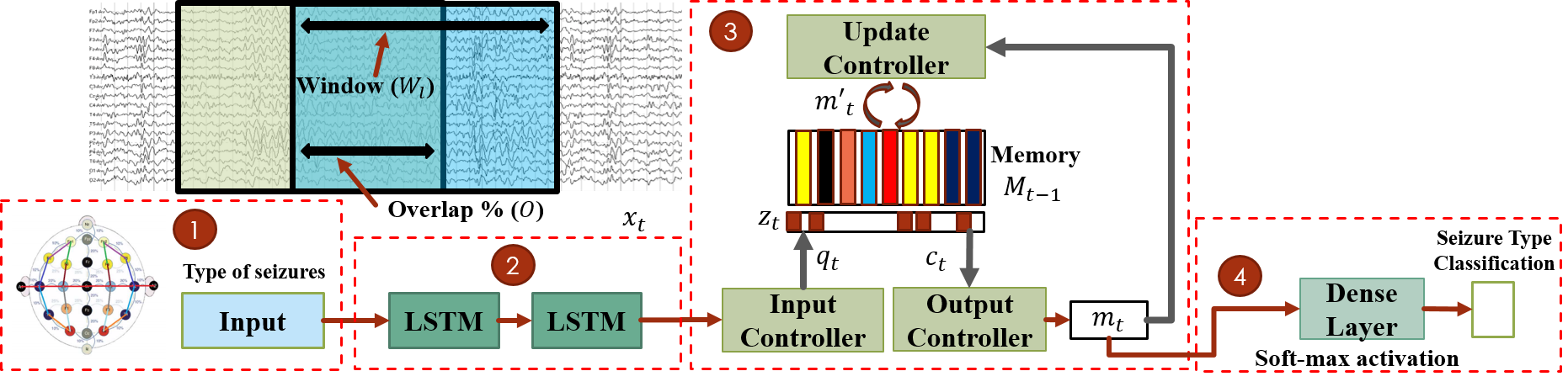} 
\end{center}
\vspace{-10pt}
   \caption{Overview of the framework proposed for classifying seizure types using sequential and neural memory networks.
   \textbf{1.} We use the TUH EEG Seizure Corpus which contains scalp EEG data from seizure recordings and a pre-processing strategy based on the fast Fourier transform. 
   \textbf{2.} We map each data sample with 2 stacked LSTMs as input to the memory model. 
   \textbf{3.} External memory model:
   The state of the memory at time instant $t-1$ is $M_{t-1}$.
   The \textit{input controller} receives the encoded hidden states $x_t$ and determines what facts within the input data to use to query the memory $q_t$. An attention score vector $z_t$ is used to quantify the similarity between the content stored in each slot of $M_{t-1}$ and the query vector $q_t$ to generate the input to the output controller.
   The \textit{output controller} regulates what results from the memory stack ($c_t$) are passed out to the memory module for the current state ($m_t$).
   The \textit{update controller} updates the memory state based on the output of the memory module and propagates it to the next time step.
   These controllers utilise a combination of fixed weights and plastic components.
   \textbf{4.} The output of the memory model is fed to a dense layer with a soft-max activation to predict each seizure type.}
\vspace{-10pt}
\label{fig:methodology}
\end{figure*}

In this paper, we propose a neural memory network (NMN), which facilitates trainable neural plasticity for robust classification of seizure types.
We compare and explore the difference between traditional deep learning techniques such as recurrent convolutional neural networks (RCNN) and our proposed framework using an external memory module.
Traditional deep learning techniques exploit short spatio-temporal relationships to model sequential data. Memory modules, on the other hand, act as a large knowledge-store, and instead of making decisions based on the current observation (input data sample), map long-term relationships across all seizure recordings.
A typical memory module~\cite{xiong2016dynamic} consists of a memory stack for information storage, a read controller to query the knowledge stored in the memory, an update controller to update the memory with new knowledge, and an output controller which controls what results are passed out from the memory.
An abstract view of these components and their interaction with the specific application of seizure classification is given in Fig.~\ref{fig:methodology}.
We compare the proposed approach with baseline algorithms for seizure detection~\cite{lin2016classification,acharya2018deep,o2018investigating,hao2018deepied,ahmedt2018deep,tsiouris2018long,thodoroff2016learning,golmohammadi2017gated} and classification~\cite{roy2019machine,asif2019seizurenet,sriraam2019convolutional}, and train all methods using supervised learning for direct comparison.

\vspace{-4pt}
\subsection{Seizure Dataset}

We use the world's largest publicly available database of 
seizure recordings, the Temple University Hospital EEG (TUH EEG)~\cite{shah2018temple} database. We focus on the subset, the TUH EEG Seizure Corpus (TUSZ, v1.4.0), which was developed to motivate research on seizure detection. 
Recordings were sampled at 250Hz 
and contain the standard channels of a 10-20 configuration.
The seizure corpus contains 2012 seizures with different lengths and eight types of seizure. 
Some seizures of the same patient are categorised with different seizure types.
Seizure recordings were annotated based on the following manifestations: electrographic, electroclinical, and clinical.
For seizure type classification experiments, we exclude only myoclonic seizures because of the small number of seizures recorded (three seizure events). The seven types of seizure selected for analysis are Focal Non-Specific Seizure (FNSZ), Generalized Non-Specific Seizure (GNSZ), Simple Partial Seizure (SPSZ), Complex Partial Seizure (CPSZ), Absence Seizure (ABSZ), Tonic Seizure (TNSZ), and Tonic Clonic Seizure (TCSZ).
The data for one seizure event consists of only the interval that contained a seizure based on the labeling reported in~\cite{shah2018temple}.
One class is defined as the combination of all seizure recordings across sessions and patients for the same seizure type.
Although we are considering here seven classes of seizure labeled in the corpus based on neurologists' reports as described in~\cite{shah2018temple}, we note that these are not clinically disjoint classes. Clinically SPSZ and CPSZ are more specific subclasses of FNSZ while ABSZ, TNSZ, and TCSZ are more specific subclasses of GNSZ. 
Thus, in cases where there was insufficient evidence to classify the type of seizure more finely, the corpus categorises the seizure event as the more general class of FNSZ or GNSZ depending on how and where it began in the brain.

We adopt the preprocessed version of TUSZ known as the IBM TUSZ pre-processed data (v1.0.0, method \#1)~\cite{roy2019machine}. 
This work used the temporal central parasagittal montage (TCP)~\cite{shah2017optimizing} of 20 selected pair channels as the input.
%
In this preprocessing method, the authors applied fast fourier transform (FFT) to each fixed-length window $W_l$ ($W_l = 1$ second and $f_{max} = 24$ frequency bands) with $O$ seconds overlap ($0.75W_l$) across all EEG recording channels, as is illustrated in Fig.~\ref{fig:methodology}.
The transformed data of all channels in one time window constitutes one data sample. Thus the task here is to perform classification based on 1 second of EEG data.
The number of data sample per seizure type corresponds to the total number of windows all seizures across all patients.
The input shape representation to train and test the model to classify seizure types is defined by [\#data sample, \#channels, \#frequency bands]. 
We adopt this input data to compare the performance of our framework with baseline results using traditional machine learning and deep learning techniques.
Table~\ref{table:seizure type} summarises the total number of seizures, patients and data samples available for each seizure type. 

\begin{table}[!t]
	\centering
		\caption{Total count for seizures and patients per seizure type.}
		\vspace{-3pt}
		\resizebox{0.46\textwidth}{!}{%
		\begin{tabular}{p{5.2cm} c c c}
				\toprule
				\textbf{Seizure Type} & \textbf{Seizures} & \textbf{Patients} & \textbf{Data sample} \\
				\midrule
				1. Focal Non-specific Seizure (FNSZ)
				& 992 & 108 & 292,725\\
				2. Generalized Non-Specific Seizure (GNSZ)
				& 415 & 44 & 137,033\\
				3. Simple Partial Seizure (SPSZ)
				& 44 & 2 & 6,028\\
				4. Complex Partial Seizure (CPSZ)
				& 342 & 34 & 132,200\\
				5. Absence Seizure (ABSZ)
				& 99 & 12 & 3,087\\
				6. Tonic Seizure (TNSZ)
				& 67 & 2 & 4,888\\
				7. Tonic Clonic Seizure (TCSZ)		 
				& 50 & 11 & 22,524\\
				\bottomrule
				\label{table:seizure type}
		\end{tabular}}
	\vspace{-30pt}
\end{table}

\vspace{-4pt}
\subsection{Traditional deep learning methods and baseline models}

Deep learning has revolutionised many medical applications and with the increasing availability of EEG datasets, these algorithms have been applied to quantify information regarding seizures~\cite{aristizabal2019understanding,ahmedt2017automated}.
We aim to adapt well-known classical deep learning structures from related domains and evaluate them for the specific task of seizure type classification.
While the objective of seizure detection is to classify the input data into two classes (a seizure class and a non-seizure class), seizure type classification aims to identify different types of epileptic seizures; \ie~it is a multi-class classification task. As such, methods initially proposed for seizure detection can be applied to the task of seizure type classification.
%

As it is not possible to consider all existing methods for seizure detection and classification in our study, we adopt only the most significant approaches based on their overall precision; the compatibility of their input data (\eg~preprocessing, image-based EEG) with the IBM TUSZ pre-processed data~\cite{roy2019machine}, and level of documentation provided by the original authors to ensure full and correct reproducibility of each model.

\subsubsection{Techniques adapted from the seizure detection domain}
The following methods are adapted to the task of seizure classification and are evaluated in this paper:
\begin{itemize}

\item Stacked auto-encoders (SAE): 
SAE are an unsupervised learning technique composed of multiple sparse autoencoders~\cite{larochelle2009exploring}. An auto-encoder consists of two parts, an encoder and a decoder. The encoder is used to map the input data to a hidden representation, and decoder is used to reconstruct input data from the hidden representation. SAE based approaches have been used by~\cite{lin2016classification,golmohammadi2019automatic}.

\item Convolutional neural networks (CNNs):
A CNN consists of multiple stacked layers of different types: convolutional layers, nonlinear layers, and pooling layers, followed by fully connected layers. Compared with traditional feed-forward neural networks, CNNs exploit spatial locality by enforcing local connectivity and parameter sharing~\cite{lecun2015deep}.
The purpose of pooling is to achieve invariance to small local distortions and reduce the dimensionality of the feature space~\cite{lecun2015deep}. 
The differences in the architecture of various proposed CNNs is due to the number of layers included in the framework, and layer parameters. Some methods fine-tune a well-known architecture such as VGG, ResNet, while other design their own deep or shallow network.
Methods including ~\cite{page2016wearable,antoniades2016deep,acharya2018deep,ullah2018automated,o2018investigating,hao2018deepied,wei2019automatic} are examples of this approach.

\item Recurrent neural networks (RNNs):
RNNs introduce the notion of time into a deep learning model by including recurrent edges that span adjacent time steps~\cite{lipton2015critical}. RNNs are termed recurrent as they perform the same task for every element of a sequence, with the output being dependant on the previous computations.
LSTMs~\cite{greff2016lstm} were proposed to provide more flexibility to RNNs by employing an external memory, termed the cell state to deal with the vanishing gradient problem. Three logic gates are also introduced to adjust this external memory and internal memory. GRUs~\cite{cho2014learning} are a variant of LSTMs which combine the forget and input gates making the model simpler. 
RNNs are used by papers including~\cite{ahmedt2018deep,tsiouris2018long}.

\item Hybrid networks:
Hybrid or cascaded networks such as recurrent convolutional neural networks (RCNN) are used to better exploit variable-length sequential data~\cite{bashivan2015learning}, to extract spatio-temporal features and classify through an end-to-end deep learning model~\cite{donahue2015long}.
In this scenario, RCNN denotes a number of convolution layers followed by stacked recurrent units (LSTMs or GRUs). 
Such methods have been proposed in~\cite{thodoroff2016learning,golmohammadi2017gated,shah2017optimizing}.
\end{itemize}

Given the dynamic nature of EEG data, RCNNs appear to be a reasonable choice for modeling the temporal evolution of brain activity.
Therefore, we have designed a shallow RCNN based on architectures used for seizure detection with video recordings as an input~\cite{aristizabal2019understanding,ahmedt2019aberrant}. We aim to demonstrate that shallow architectures are capable of reaching similar results to more complex traditional deep learning models.
Through extensive experiments, the design of the network architecture that shows the best performance consists of 
1) CNN: two convolutional layers (32 kernels of size $3 \times 3$) stacked together followed by one max-pooling layer (size $2 \times 2$) and a fully-connected layer (512 nodes);
2) LSTM: two LSTM layers each with 128 cells followed by a densely connected layer with a softmax activation layer.

\subsubsection{Baseline methods for seizure type classification}
The following methods are compared directly with the proposed approach:
\begin{itemize}
\item Traditional machine learning techniques: K-Nearest Neighbors, SGD, XGBoost, and AdaBoost classifiers were proposed in~\cite{roy2019machine}.
\item Traditional CNNs: a residual network ResNet50 was retrained to perform classification in~\cite{roy2019machine}.
Three pretrained models, AlexNet, VGG16 and VGG19, were used in~\cite{sriraam2019convolutional} to solve the classification problem. However, an additional class of non-seizure events was included in this publication.
\item SeizureNet~\cite{asif2019seizurenet}: the authors proposed two sub-networks, a deep convolutional network (multiple bottleneck convolutions interconnected through dense connections) and a classification network.
\end{itemize}

\vspace{-5pt}
\subsection{Neural memory networks and neural plasticity}

The design of the network architecture for the task of seizure type classification from EEG recordings is displayed in Fig.~\ref{fig:methodology}. 
This approach aims to update the external memory model (a memory stack for information storage) with new information from each data sample, and as such the memory learns to store distinctive characteristics from each seizure type across patients.
First, for modelling short-term relationships within the data sample we use LSTMs. To extract the relevant attributes through long-term dependencies (across seizures and patients), we employ the proposed neural memory architecture. 
The seizure classification output is generated using a dense layer with softmax classification.

The neural memory architecture is composed of a memory stack $M$, with $l$ memory slots each with an embedding size $k$ ($M \in \mathbb{R} ^{l \times k}$), and its respective input, output and update controllers. Each of these controllers is composed of an LSTM cell following~\cite{munkhdalai2017neural,fernando2018tree}. 
The input controller passes the encoded hidden state from the stacked LSTMs, $x_t$, at time instant $t$ and generates a vector, $q_t$, to retrieve the salient information from the stored knowledge in the memory. 
We generate an attention score vector $z_t$ to quantify the similarity between $q_t$ and the content of each slot of $M_{t-1}$.
Then, the output controller can retrieve the memory output, $m_t$, for the current state. We pass this resultant embedding through an update controller to generate an updated vector $m^{'}_t$, which is used to update the memory and propagate it to the next time step. We update the content of each memory slot based on the informativeness reflected in the score vector~\cite{tharindu2019TNNLS}.
We define the input, output and update operations such that,
\begin{equation}
q_t = f_{\text{input}}^{\text{LSTM}} (x_t) ,
\end{equation}
\begin{equation}
z_t = \text{softmax}(q_t^T M_{t-1}) ,
\end{equation}
\begin{equation}
m_t = z_tM_{t-1} ,
\end{equation}
\begin{equation}
m^{'}_t = f_{\text{update}}^{\text{LSTM}} (m_t) ,
\end{equation}
\begin{equation}
M_t = M_{t-1}(I-(z_t \otimes e_k)^T) + (m^{'}_t \otimes e_l)(z_t \otimes e_k)^T ,
\end{equation}
where $I$ is a matrix of ones, $e_l$ and $e_k$ are vectors of ones and $\otimes$ denotes the outer vector product which duplicates its left vector $l$ or $k$ times to form a matrix.
Ideally, we expect that the memory output $m_t$ should capture salient information from both the input and stored history that can be used to estimate each type of seizure.

Inspired by the success of~\cite{miconi2018differentiable} in demonstrating how neural plasticity can be optimized by gradient descent in recurrent networks, we adopt neural plasticity to enhance the memory access mechanisms in the memory model.
To perform the injection of plasticity for memory components, we adopt the formulation of the Hebbian rule for its flexibility and simplicity (``neurons that fire together, wire together'')~\cite{miconi2018differentiable}.
We define a fixed component (a traditional connection weight $w$) and a plastic component for each pair of neurons $i$ and $j$, where the plastic component is stored in a Hebbian trace $\text{Hebb}$, which evolves over time based on the inputs and outputs. 
The Hebbian trace is simply an average of the product of pre- and post-synaptic activity. Thus, the network equation for the output $x_t$ of neuron $j$ are:
\begin{equation}
x^j_t = \tanh ( \sum_{i \in inputs} [w^{i,j}x^i_{t-1} + \alpha^{i,j}\text{Hebb}^{i,j}_tx^i_{t-1}]  ) ,
\end{equation}
\begin{equation}
\text{Hebb}^{i,j}_{t+1}= \text{Hebb}^{i,j}_t \eta x^j_t( x^i_{t-1} - x^j_t \text{Hebb}^{i,j}_t ) ,
\end{equation}

Here $\alpha$ controls the contribution from fixed and plastic terms of a particular weight connection, and $\eta$ is the learning rate of plastic components.
Thus, we replace the component of the controllers to produce a plastic neural memory such that,
\begin{equation}
q_t = \tanh ( \sum_{\forall i,j \in k} [\dot{w}^{i,j}x^i_{t-1} + \dot{\alpha}^{i,j}\dot{\text{Hebb}}^{i,j}_tx^i_{t-1}]  )  ,
\end{equation}
\begin{equation}
c_t = z_tM_{t-1} ,
\end{equation}
\begin{equation}
m_t = \tanh ( \sum_{\forall i,j \in k} [\hat{w}^{i,j}x^i_{t-1} + \hat{\alpha}^{i,j}\hat{\text{Hebb}}^{i,j}_tc^i_{t-1}]  )  ,
\end{equation}
\begin{equation}
m^{'}_t = \tanh ( \sum_{\forall i,j \in k} [\tilde{w}^{i,j}x^i_{t-1} + \tilde{\alpha}^{i,j}\tilde{\text{Hebb}}^{i,j}_tm^i_{t-1}]  )  ,
\end{equation}
Further technical information on neural memory networks and plasticity can be found in~\cite{munkhdalai2017neural,fernando2018task,fernando2018tree,tharindu2019TNNLS}.

\section{EVALUATION}

\subsection{Experimental setup}

All models were assessed through a 5-fold cross validation (CV) strategy to ensure that the data for hyperparameter tuning, and the data to test the algorithm were disjoint. 
For each fold, the data sample of each seizure type are randomly split into 60\% for training, 20\% for validation and 20\% for test. 
%
We used a weighted-F1 score to measure performance as this is a multiclass classification with highly uneven class distribution. The weighted F1 score is calculated as follows,
\begin{equation}
\text{Weighted} \; F_1 = \sum_{i=1}^{7} \dfrac{2 \times \text{precision}_i \times \text{recall}_i} {\text{precision}_i + \text{recall}_i} \times w_i,
\end{equation}
where $w_i$ is the weight of the $i-th$ class depending on the number of positive examples in that class.

For each traditional deep learning models adapted for the task of seizure type classification, we follow the specifications provided by each author to train the architecture by optimizing the categorical cross-entropy loss. 
Our proposed shallow RCNN model was also trained by optimizing the categorical cross-entropy loss. We used the Adam optimizer~\cite{kingma2014adam} with a learning rate of $10^{-3}$, and decay rates for the first and second moments of 0.9 and 0.999 respectively. 
For regularization, we employed dropout with a probability of 50\% in the fully connected layer. Batch-size was set to 32. We trained the model over 150 epochs using the default initialization parameters from Keras~\cite{chollet2017keras}.

For the proposed plastic NMN model, we also adopt the Adam optimizer and categorical cross-entropy loss and train for 50 epochs. Hyper-parameters $k=80$ (hidden state dimension), $l=25$ (memory length), and $\eta=0.5$ (learning rate of plasticity) were evaluated experimentally, and were chosen as they provide the best accuracy on the validation set.

\vspace{-8pt}
\subsection{Classification of seizure type}

\begin{figure}[t]
\begin{center}
\includegraphics[width=0.6\linewidth]{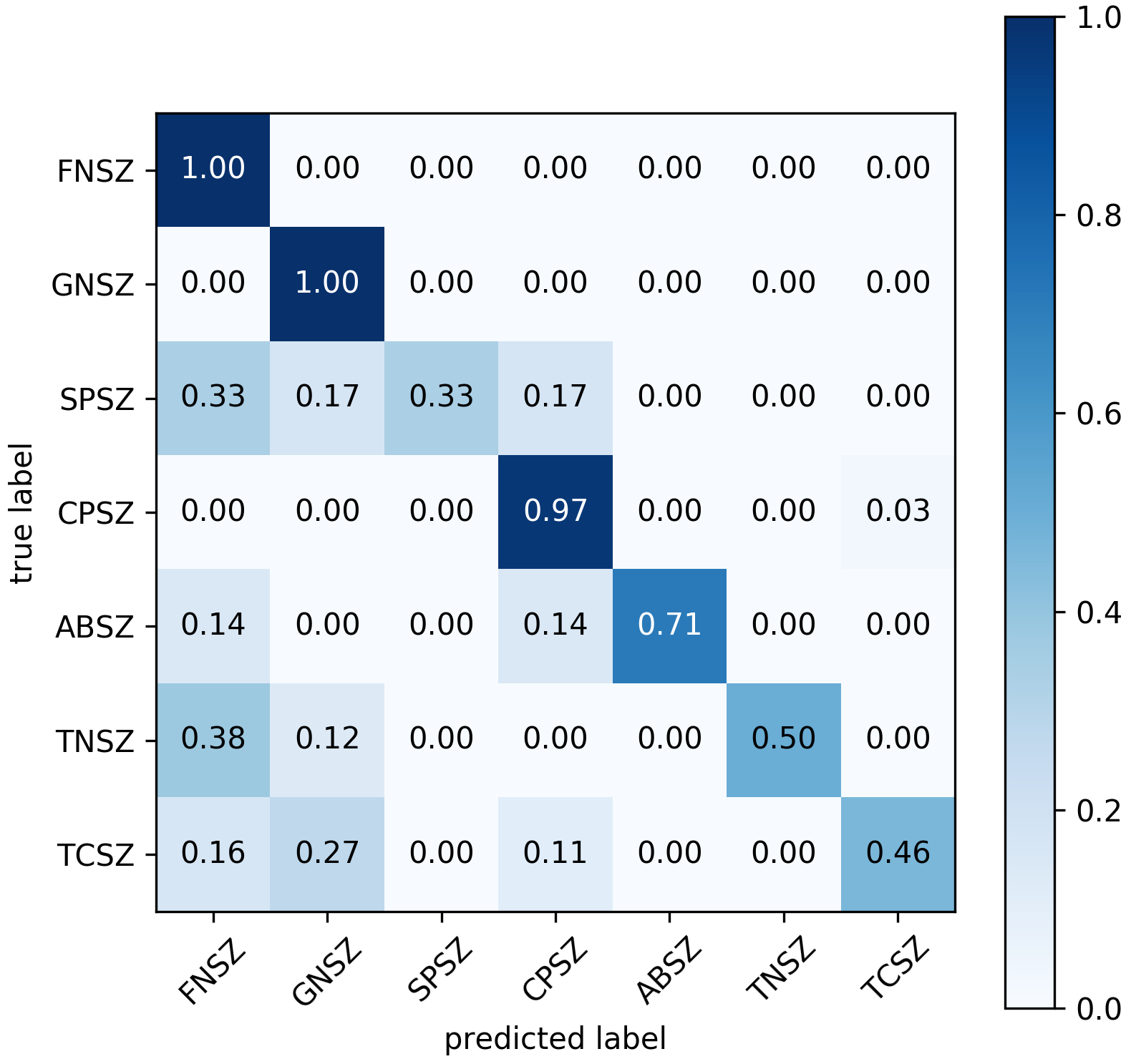} 
\end{center}
\vspace{-6pt}
   \caption{Normalized confusion matrices for seizure type classification on the TUH EEG Seizure Corpus for the proposed Plastic NMN model.} 
\vspace{-12pt}
\label{fig:cmatrices}
\end{figure}

Table~\ref{table:results} summarizes the results of seizure type classification on the TUH EEG Seizure corpus with the IBM TUSZ pre-processed data using our proposed framework, along with the baseline methods and the adapted methodologies from the seizure detection domain.
It is evident that through the utilization of the proposed external memory model via augmented read and write mechanisms with plasticity, we were able to achieve superior classification results.
Fig.~\ref{fig:cmatrices} shows the normalized confusion matrices of the seven types of seizure for the proposed Plastic NMN method.
In this confusion matrix, we can identify that the most difficult seizure types to discriminate are those which have a small number of seizure recordings available for training and testing (\ie SPSZ, ABSZ, TNSZ and TCSZ).

\begin{figure}[!t]
\begin{center}
\includegraphics[width=0.66\linewidth]{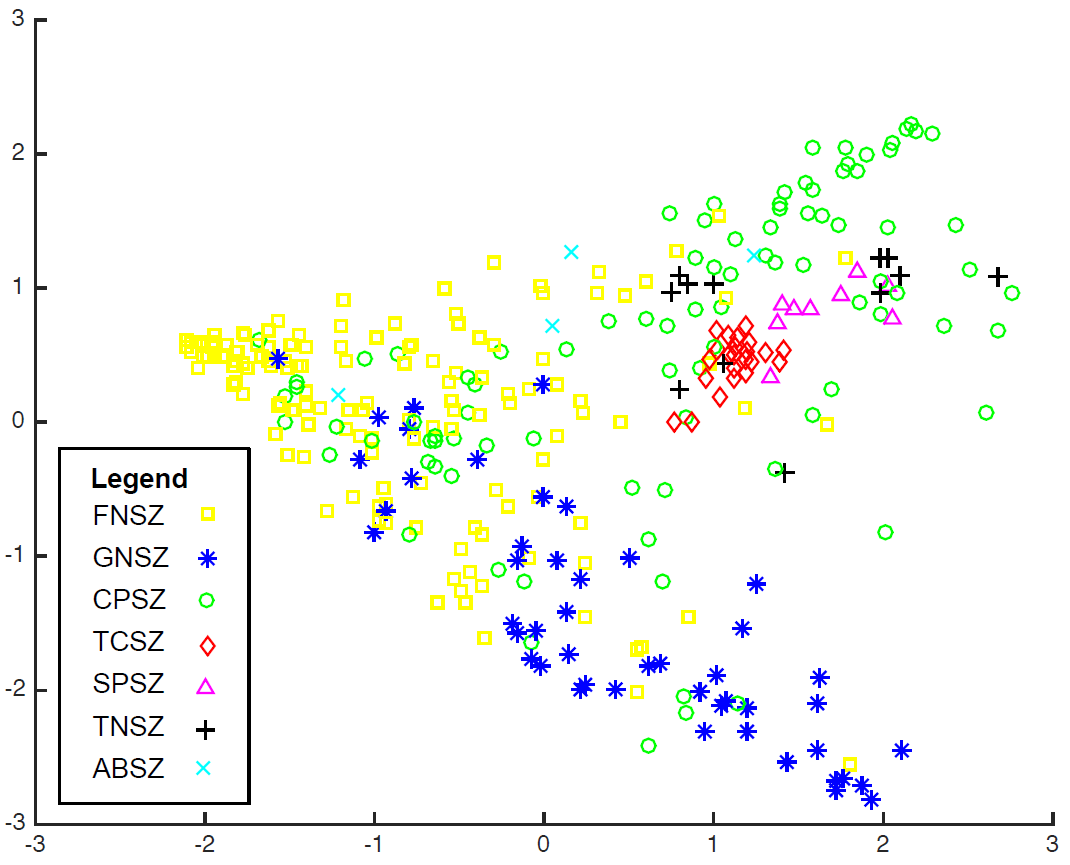} \end{center}
\vspace{-4pt}
   \caption{2D illustration of extracted embeddings from the CNN-LSTM model for randomly selected 500 samples from the test set.}
\vspace{-6pt}
\label{fig:clustersRCNN}
\end{figure}

To qualitatively illustrate the significance of the salient information and what the model has learned in terms of the model activations, we randomly sample 500 inputs from the test set and apply PCA~\cite{jolliffe2011principal} and plot the top two components in 2D. The embeddings are extracted from the last LSTM layer in the RCNN model and from the external memory in the plastic NMN.
Fig.~\ref{fig:clustersRCNN} and Fig.~\ref{fig:clustersNMN} depict the resultant plot where each seizure type is indicated based on the ground truth class identity.
We observe clear separation between the seven type of seizures using the memory embeddings compared to the features learnt by our proposed RCNN model or SeizureNet~\cite{asif2019seizurenet}.
This clearly demonstrates that the resultant sparse vectors are sufficient to discriminate between classes with simple classifiers.


\begin{table}[!t]
\caption{Cross-validation performance of classifying seizure type}
\vspace{-5pt}
\centering
\label{table:results}
\resizebox{0.32\textwidth}{!}{%
\begin{tabular}{>{\raggedleft\arraybackslash}p{4.5cm}
>{\centering\arraybackslash}p{2.5cm}
}
\toprule
\textbf{Baseline methods}  & \textbf{Weighted-F1 score} \\ 
\midrule
Adaboost~\cite{roy2019machine} & 0.509 \\
SGD~\cite{roy2019machine} & 0.649 \\
XGBoost~\cite{roy2019machine} & 0.782 \\
KNN~\cite{roy2019machine} & 0.884 \\
CNN (ResNet50)~\cite{roy2019machine} & 0.723 \\
CNN (AlexNet)~\cite{sriraam2019convolutional} & 0.802\\
SeizureNet~\cite{asif2019seizurenet} & 0.900 \\
\midrule
\textbf{Baseline from adapted methods}  & \textbf{Weighted-F1 score} \\ 
\midrule
SAE (based on~\cite{lin2016classification}) & 0.675 \\
CNN (based on~\cite{acharya2018deep}) &  0.716 \\
CNN (based on ~\cite{o2018investigating}) & 0.826 \\
CNN (based on~\cite{hao2018deepied}) & 0.901 \\
LSTM (based on~\cite{ahmedt2018deep}) &  0.692 \\
LSTM (based on~\cite{tsiouris2018long}) & 0.701 \\
CNN-LSTM (based on~\cite{thodoroff2016learning}) & 0.795 \\
CNN-LSTM (based on~\cite{golmohammadi2017gated}) & 0.831 \\
CNN-LSTM (this work) & 0.824 \\
\midrule
\textbf{Proposed framework}  & \textbf{Weighted-F1 score} \\ 
\midrule
Plastic NMN (this work) & 0.945 \\
\bottomrule
\end{tabular}}
\vspace{-10pt}
\end{table}

\vspace{-8pt}
\subsection{Discussion}

Several studies have demonstrated that machine learning models, specifically deep learning networks, can successfully detect and/or predict the onset of seizures from scalp and intracranial EEG.
Although such models may be useful in identifying biomarkers of an existing epileptic condition, they are rarely of use for discriminating between different type of seizures. 
In this paper, we have evaluated traditional deep learning methods proposed in the epilepsy domain for cross-patient seizure type classification, and we have improved on existing reported results by presenting a neural memory network based framework.

We note that RCNNs have reached better performance than models based on CNNs or RNNs alone for the task of seizure type classification, which is similar to their relative performance reported for the seizure detection task.
Given the inherent temporal structure of EEGs, we expected that recurrent networks would be more widely employed than models that do not consider time dependencies. However, almost half of the models proposed in the epilepsy domain have used CNNs. This observation supports recent discussions regarding the effectiveness of CNNs for processing time series data~\cite{bai2018empirical}.
Another finding of our study of baseline models is that the shallow RCNN proposed performed as well as deep CNNs models.
This supports other research that has preferred shallow networks for analysing EEG data. 
Schirrmeister \etal~\cite{schirrmeister2017deep} focused on this aspect, comparing the performance of architectures with different depths and structures. The authors showed that shallower fully convolutional models outperformed their deeper counterparts. However, we note that hyperparameter tuning of baseline models may be key to using deeper architectures with physiological recordings.

The potential of recurrent neural networks to handle sequence information was evident in the experimental results. However, it is essential to consider historic behaviour over the full length of seizures, and map long-term dependencies between seizures to generate more precise classification. 
The process of capturing seizure behaviour is highly complex because of the increased heterogeneity of participants and the temporal evolution during epileptic seizures.
Analyzing dynamic changes during a seizure is a major aspect of epilepsy patient assessment.
%
Even though an RNN model has the ability to capture temporal information, it considers only the relationships within the current sequence due to the internal memory structure, making accurate long-term prediction intractable.
RNNs such as LSTMS or GRUs exhibit one common limitation related to their storage capacity because their internal state is modified, heavily or slightly, at each computation step.
By incorporating a neural memory network, we are able to increase the model's storage capacity without having to increase the size of the model; as demonstrated by~\cite{munkhdalai2017neural}, who compared using neural memory networks to map long-term dependencies among the stored facts with LSTMs which map dependencies within the input sequence.

The memory network proposed in this paper is capable of capturing both short-term
(within each data sample) as well as long-term (across the entire collection of data samples) relationships to predict a seizure type (\ie~long-term memory and working memory).
Therefore, our proposed system eliminates the deficiencies of current baseline models in epilepsy classification which only consider within-sequence relationships.
An additional benefit of the implemented memory network is that we have introduced the concept of synaptic plasticity through the read and write operation of the learnable controllers.
We apply local plasticity rules (Hebbian trace) to update feed-forward synaptic weights following feedback projection.
The plastic nature of the memory access mechanisms in the neural memory model allows our system to provide a varying level of attention to the stored information, \ie~the plastic network acts as a content-addressable memory.

\begin{figure}[!t]
\begin{center}
\includegraphics[width=0.66\linewidth]{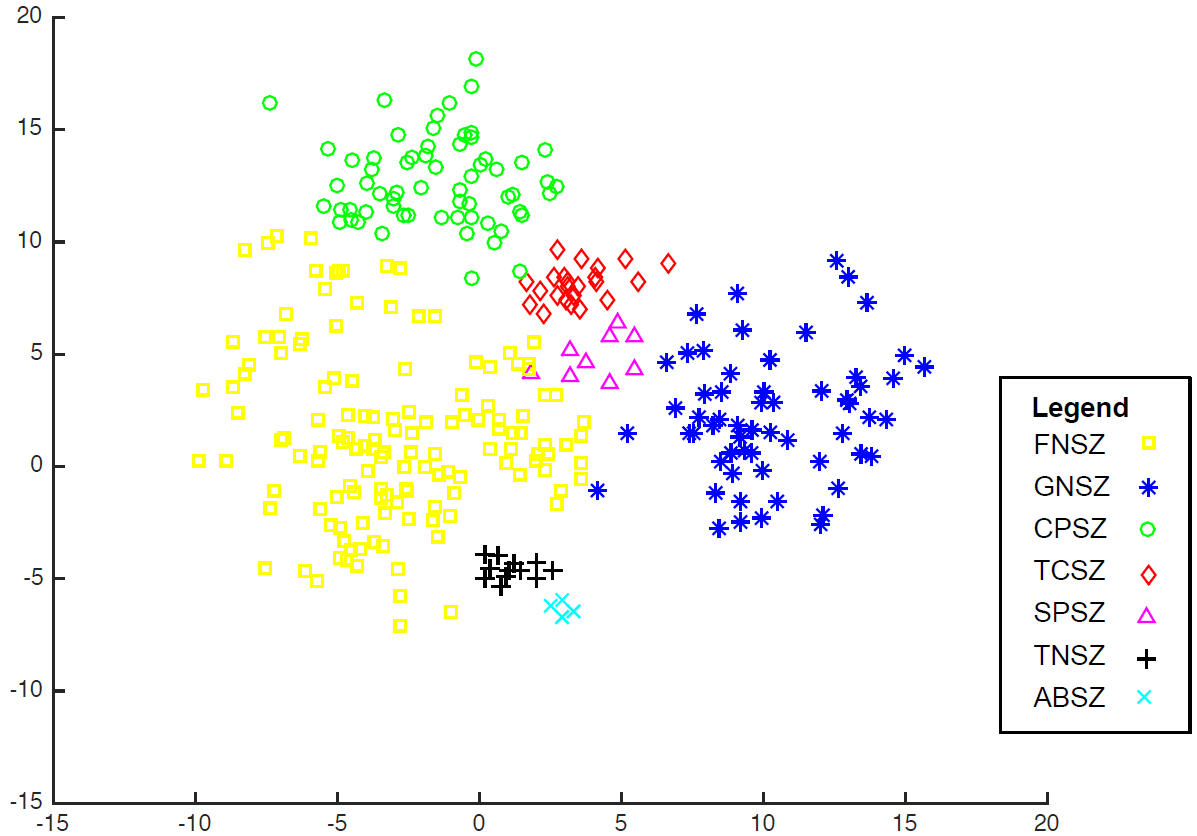} 
\end{center}
\vspace{-4pt}
   \caption{2D illustration of extracted memory embeddings from the Plastic NMN for randomly selected 500 samples from the test set.}
\vspace{-9pt}
\label{fig:clustersNMN}
\end{figure}

To allow comparison with baseline methods~\cite{roy2019machine,asif2019seizurenet,sriraam2019convolutional} we defined the classification task here similarly: to separate seven classes of seizure labeled in the Corpus. As noted above, these classes are not actually clinically disjoint, but form a hierarchy. This semantic structure is not exploited by the present method. Where for example in Fig.~\ref{fig:clustersNMN} more specific seizure classes such as ABSZ are readily separated from the more general class GNSZ, this may indicate overfitting due to the small number of distinct patients for some seizure classes in the Corpus. This shows the value of continuing to expand the seizure corpus with more patients for future work.

\section{CONCLUSIONS}
\vspace{-3pt}

This paper presents a deep learning based framework which consists of a neural memory network with neural plasticity for EEG-based seizure type classification.
A brief overview of commonly used deep learning approaches in the epilepsy domain is also presented. 
The proposed approach is capable of modelling long-term relationships which enables the model to learn rich and highly discriminative features for seizure type classification.
With increasing computational capabilities and the collection of larger datasets, clinicians and researchers will increasingly benefit from the significant progress already made in their application to epilepsy.
An accurate classification of seizures along with neuroimaging and behavioural analsyis are one step towards more accurate prognosis.
In future, we plan to investigate the introduction of the memory component to map relationships directly from raw intracranial EEG recordings without a preprocessing phase, \ie.~without extracting information contained in the frequency transform of the time-series EEG.






\bibliographystyle{IEEEtran}
\bibliography{egbib}

\vspace{0.5cm}

\end{document}